\documentclass[showpacs,preprintnumbers,amsmath,amssymb]{revtex4}
\usepackage{amsmath,amssymb,graphics,epsfig,subfigure}
\usepackage{color}

\begin{document}
\renewcommand{\baselinestretch}{1.3}

\title{Analytical and exact critical phenomena of $d$-dimensional singly spinning Kerr-AdS black holes}

\author{Shao-Wen Wei \footnote{Corresponding author. weishw@lzu.edu.cn},
        Peng Cheng \footnote{pcheng14@lzu.edu.cn},
        Yu-Xiao Liu \footnote{liuyx@lzu.edu.cn}}

\affiliation{Institute of Theoretical Physics, Lanzhou University, Lanzhou 730000, People's Republic of China}

\begin{abstract}
In the extended phase space, the $d$-dimensional singly spinning Kerr-anti-de Sitter black holes exhibit the van der Waals phase transition and reentrant phase transition. Since the black hole system is a single-characteristic-parameter system, we show that the form of the critical point can be uniquely determined by the dimensional analysis. When $d=4$, we get the analytical critical point. The coexistence curve and phase diagram are obtained. The result shows that the fitting form of the coexistence curve in the reduced parameter space is independent of the angular momentum. When $d=5$---$9$, the exact critical points are numerically solved. It demonstrates that when $d\geq6$, there are two critical points. However, the small one does not participate in the phase transition. Moreover, the exact critical reentrant phase transition points are also obtained. All the critical points are obtained without any approximation.
\end{abstract}



\maketitle

\section{Introduction}
\label{secIntroduction}

The subject of black hole thermodynamics continues to bean important topic in gravitational physics. Inspired in particular by the AdS/CFT correspondence \cite{Maldacena,Gubser,Witten}, the thermodynamical properties and phase transitions of the black holes in anti-de Sitter (AdS) space have gained much attention. Utilizing the AdS/CFT correspondence, the Hawking-Page phase transition between a stable large black hole and a thermal gas \cite{Hawking} was explained as the confinement/deconfinement phase transition of a gauge field \cite{Hawking,Witten2}.

The study was also generalized to the charged AdS black hole. In the grand canonical ensemble with fixed electric potential, there also exists the Hawking-Page phase transition. However, in the canonical ensemble with fixed charge, $1/T$---$r_{h}$ displays an oscillatory behavior and the free energy exhibits a swallow tail behavior, which confirms a small/large black hole phase transition reminiscent of a liquid/gas transition of the van der Waals (vdW) fluid \cite{Chamblin,Chamblin2,Roychowdhury,Banerjee}. At the critical point, the phase transition will become a second-order one. Similar critical phenomena can also be found for the Kerr-Newman-AdS black hole \cite{Caldarelli}. Moreover, the critical phenomena were also found in the $Q$-$\Phi$ diagram for the charged AdS black hole \cite{Wu,Shen}, and for other AdS black holes \cite{Niu,Tsai,Wei}.

Although the small/large black hole phase transition behaves like a vdW fluid, the precise analogy was not established until recently by Kubiznak and Mann in Ref. \cite{Kubiznak} in the extended phase space, where the cosmological constant was treated as the thermodynamic pressure and its conjugate quantity as the thermodynamic volume of the AdS black hole \cite{Kastor,Dolan00,Cvetic}. This rejuvenated the study of the AdS black hole thermodynamics, including the $P$-$v$ criticality, phase structure, and so on. Many works demonstrated that such critical phenomena are presented in charged AdS black hole systems \cite{Hendi,Chen,ZhaoZhao,Cai,XuXu,Mo,zou,MoLiu,Moliu,Molili,Zou2,ZhaoZhang,Mirza,Kostouki,DolanBP,
Kamrani,Wei3,Spallucci,Smailagic,XuXuzhao,Cai2,Cao0,MajhiMajhi}. They have the same critical exponents and scaling laws, as well as the similar $P$-$v$ oscillatory behaviors and coexistence curves. In analogy to a vdW fluid, we recently introduced the number density of black hole molecules to measure the microscopic degrees of freedom of black holes \cite{Wei4}. It was found that the number density suffers a sudden change accompanied by a latent heat when the black hole system crosses the small-large black hole coexistence curve. However, when the system passes the critical point, it encounters a second-order phase transition with a vanishing latent heat due to the continuous change of the number density. Therefore, the difference between the number densities of the small and large black holes provides us with an order parameter to describe the small/large black hole phase transition. The system is also found to possess a maximum number density $n/n_{c}=2.44$, which is slightly smaller than $n/n_{c}=3$ of the vdW fluid. These results further confirm that the small-large black hole phase transition is of vdW type. Moreover, multicritical phenomena---such as the reentrant phase transition and triple point---were also observed in Refs. \cite{Gunasekaran,Wei2,Frassino,Sherkatghanad,Panahiyan}.

In the extended phase space, the rotating AdS black holes have richer phase structure. Through the Gibbs free energy, the four-dimensional Kerr-AdS black hole was found to possess a small/large black hole phase transition \cite{Gunasekaran}, which is similar to the charged AdS black hole. The five-dimensional singly rotating Kerr-AdS black hole also has such a phenomenon \cite{Altamirano3}. However, when the dimension $d$ increases or the multiply rotating parameters are presented, the case becomes much more exciting. The authors of Ref. \cite{Altamirano} first found the reentrant large/small/large black hole phase transition in the $d=6$-dimensional singly spinning Kerr-AdS black hole system. For the $d=6$-dimensional multiply spinning Kerr-AdS black hole with two fixed angular momenta $J_{1}$ and $J_{2}$, the system has a triple point and solid/liquid/gas analogue phase transition in some parameter ranges \cite{AltamiranoKubiznak}, and more than one critical point was found. A similar phase structure can also be found for higher-dimensional or multiply-rotating-parameter Kerr-AdS black holes. These exciting results imply that the AdS black hole system behaves like a multicomponent fluid system.

Among the study of the critical phenomena of the rotating AdS black hole, one important issue is to find the exact critical point. For a specific value of the angular momentum $J$, one can numerically solve the critical point. However, the specific form of the point is very hard to obtain. In Refs. \cite{Gunasekaran,Altamirano3}, the authors first expanded the state equation and all quantities in the small-$J$ limit and obtained an approximate critical point for the $d$-dimensional singly rotating Kerr-AdS black hole. As is known, the appearance of the reentrant phase transition requires more than one critical point. So there still exists another critical point that has not been found for $d\geq6$. Another question is how the critical points behave at large $J$.

The purpose of this paper is to find the exact critical points for the $d$-dimensional singly spinning Kerr-AdS black holes. The paper is organized as follows. In Sec. \ref{Classification}, we divide the thermodynamic quantities into two classes according to their different properties. Based on this, we get the explicit form of the critical point using the dimensional analysis. In Sec. \ref{Maxwell}, we discuss the generalized Maxwell equal-area laws in the extended phase space for the Kerr-AdS black hole, and give three critical conditions, which can be used to determine the critical points. In Sec. \ref{KerrAdS}, the analytical critical point for the four-dimensional Kerr-AdS black hole is derived. The phase diagram and the coexistence curve are obtained. Moreover, the exact critical points for $d=5$---9 cases are derived. It is also found that there exist two critical points for $d\geq6$. Comparing the approximate critical points obtained in the small-$J$ limit \cite{Gunasekaran,Altamirano3} with our exact ones, we find that the approximate critical points are not only accurate for small $J$, but they are also valid for large $J$. The exact critical reentrant phase transition points (CRPs) are derived in Sec. \ref{reentrant}. Finally, we summarize and discuss our results in Sec. \ref{Conclusion}.

\section{Classification of thermodynamic quantities}
\label{Classification}

Before investigating the critical phenomena, we would like to give a simple classification of the thermodynamic quantities, and discuss the relation between the classification and the critical point.

Let us start with the equation of state of a vdW fluid,
\begin{equation}
 T=\bigg(P+\frac{a}{v^{2}}\bigg)(v-b),
\end{equation}
where $v=V/N$ is the specific volume with $V$ and $N$ being the volume and total number of the molecules for the fluid system. The parameters $a$ and $b$ come from the corrections of attraction between the molecules and the nonzero size of the molecules. For different fluids, $a$ and $b$ take different values. From this equation, we have five parameters: $P$, $V$, $T$, $a$, and $b$. In general, we can divide them into two classes. The first class consists of \emph{universal parameters} ---such as the pressure $P$, volume $V$, and temperature $T$---which describe the universal properties of an ordinary thermodynamic system. The second class consists of the \emph{characteristic parameters}---such as $a$ and $b$---which describe the characteristic properties of a thermodynamic system.

With such classifications, we can further discuss the critical point and the coexistence curve. After examining the concept of the critical point, we note that the critical point can be interpreted as the relation between the universal and characteristic parameters. For example, one has the following critical point for the vdW fluid:
\begin{equation}
 P_{c}=P_c(a, b),\quad V_{c}=V_c(a, b),\quad T_{c}=T_c(a, b).\label{cr}
\end{equation}
And the $P$-$T$ coexistence curve is
\begin{equation}
 P=f(a, b, T),\label{coe}
\end{equation}
where $f(a, b, T)$ is a function of $a$, $b$, and $T$. It seems that such an analysis does not provides us with a new result for the phase transition. However, if a system has only one characteristic parameter, what will happen? For example, the $d$-dimensional charged AdS black hole and singly spinning Kerr-AdS black hole are both single-characteristic-parameter systems.

The critical point and coexistence curve for $d$-dimensional charged AdS black hole are already known. We will first take it as a simple example, and then apply the analysis to the rotating Kerr-AdS black hole. We first list the length dimensions of the relevant parameters in units of $c=G_{d}=\hbar=k_{B}=1$,
\begin{eqnarray}
 [P]&=& L^{-2},\quad [T]=L^{-1}, \quad [V]=L^{d-1},\nonumber\\
 ~[v]&=& L,\quad [J]=L^{d-2}, \quad [Q]=L^{d-3}, \quad [S]=L^{d-2}.\label{dimension}
\end{eqnarray}
Since the charged AdS black hole is a single-characteristic-parameter system, the critical point should have the following form according to Eq. (\ref{cr}):
\begin{equation}
 P_{c}=\alpha_{1} Q^{-\frac{2}{d-3}},\quad
 T_{c}=\alpha_{2}Q^{-\frac{1}{d-3}},\quad
 S_{c}=\alpha_{3}Q^{\frac{d-2}{d-3}},\quad
 V_{c}=\alpha_{4}Q^{\frac{d-1}{d-3}},\quad
 v_{c}=\alpha_{5}Q^{\frac{1}{d-3}},\quad
 \label{chargeAdS}
\end{equation}
where $\alpha_{1}\sim\alpha_{5}$ are dimensionless constants and can be determined by the critical condition, which we will discuss in the next section. The specific values for these coefficients can be found in Ref. \cite{Gunasekaran}. The parametrization form of the coexistence curve can also be found in Ref. \cite{Wei3}, which will be in the form of Eq. (\ref{coe}) if we express them in the ordinary parameter space. Note that we obtain the form for the critical points without any calculation.

Now, we would like to turn to the singly spinning Kerr-AdS black hole. Combined with the dimensional analysis, the critical point must be in the form
\begin{equation}
 P_{c}=\beta_{1} J^{-\frac{2}{d-2}},\quad
 T_{c}=\beta_{2}J^{-\frac{1}{d-2}},\quad
 S_{c}=\beta_{3}J,\quad
 V_{c}=\beta_{4}J^{\frac{d-1}{d-2}}\quad
 v_{c}=\beta_{5}J^{\frac{1}{d-2}}, \label{criticalpoint}
\end{equation}
where $\beta_{1}\sim\beta_{5}$ are dimensionless constants. Now we have given the explicit form of the critical point without any approximation for the singly spinning Kerr-AdS black hole. From such an analysis, we see that the exact value of the critical point should exist, and the coefficients $\beta_{i}$ will be determined in the following sections.

Another byproduct is that the universal ratios $\rho_{1}=\frac{P_{c}v_{c}}{T_{c}}$ and $\rho_{2}=\frac{T_{c}V_{c}}{S_{c}}$ are dimensionless constants. For the charged AdS black hole, we have $\rho_{1}=\frac{2d-5}{4d-8}$ \cite{Gunasekaran} and $\rho_{2}=\frac{4(d-3)^{2}}{(d-1)(2d-5)\pi}$.

\section{Generalized Maxwell equal-area laws}
\label{Maxwell}

In this section, we will examine the equal-area laws for the rotating AdS black hole, and use them to show the critical conditions.

In the extended phase space, the cosmological constant is treated as a new thermodynamic variable \cite{Kubiznak,Dolan00}, i.e., the pressure
\begin{eqnarray}
 P=-\frac{\Lambda}{8\pi}=\frac{(d-1)(d-2)}{16\pi l^{2}}.
\end{eqnarray}
With such an interpretation of the cosmological constant, the first law of the black hole thermodynamics is modified as
\begin{equation}
 dH\equiv dM=TdS+VdP+\Omega dJ.
\end{equation}
Compared with that of the vdW fluid, the black hole mass $M$ should be identified with the enthalpy $H$ rather than the internal energy, and $V=(\partial_{P}H)_{S,J}$ is related to the thermodynamic volume of the system. This system has three pairs of intensive/extensive variables, i.e., $(T, S)$, $(P, V)$, and $(J, \Omega)$. The Gibbs free energy for the system reads
\begin{equation}
 G=H-TS.
\end{equation}
Differentiating it, we get
\begin{equation}
 dG=-SdT+VdP+\Omega dJ.
\end{equation}
Each point located on the coexistence curve corresponds to two states, labeled as ``A" and ``B". They are in different system phases but have the same Gibbs free energy. Then we have $dG=G_{A}-G_{B}=0$, which leads to
\begin{equation}
 -SdT+VdP+\Omega dJ=0.
\end{equation}
Integrating the above equation from ``A" to ``B", one easily obtains
\begin{equation}
 -\int_{T_{A}}^{T_{B}}SdT+\int_{P_{A}}^{P_{B}}VdP
    +\int_{J_{A}}^{J_{B}}\Omega dJ=0.\label{area}
\end{equation}
As was done in Ref. \cite{Wei3}, we can obtain the generalized equal-area laws by only varying one parameter and keeping the other fixed.

\noindent\textbf{Case I:} $J$ and $P$ fixed. Equation (\ref{area}) reduces to
\begin{equation}
\int_{T_{A}}^{T_{B}}SdT=0.
\end{equation}
Then following the simple calculation of Ref. \cite{Wei3}, we can construct the equal-area law on the $T$-$S$ oscillatory line.

\noindent\textbf{Case II:} $J$ and $T$ fixed. Equation (\ref{area}) gives
\begin{equation}
\int_{P_{A}}^{P_{B}}VdP=0.
\end{equation}
Thus, we can construct the equal-area law on the $P$-$V$ oscillatory line.

\noindent\textbf{Case III:} $P$ and $T$ fixed. We have
\begin{equation}
 \int_{J_{A}}^{J_{B}}\Omega dJ=0.
\end{equation}
The equal-area law can also be constructed in a similar way.

In summary, there are three oscillatory lines: the $T$-$S$, $P$-$V$, and $J$-$\Omega$ lines. Along these three curves, we can construct three equal-area laws. They are consistent with each other and are effective for finding the phase transition points.

It was recently found that there exists a small/large black hole phase transition of the vdW type for a rotating AdS black hole \cite{Gunasekaran}. In general, the coexistence curve shown in the $P$-$T$ diagram has a positive slope, and it terminates at a critical point as the temperature $T$ increases. The critical point can be determined by the oscillatory curves. Since there are three curves, we have three kinds of critical conditions to determine the critical point. Here we list all of them:
\begin{eqnarray}
 (\partial_{V}P)_{J,T}&=&(\partial_{V,V}P)_{J,T}=0,\label{crit1}\\
 (\partial_{S}T)_{J,P}&=&(\partial_{S,S}T)_{J,P}=0,\label{crit2}\\
 (\partial_{\Omega}J)_{P,T}&=&(\partial_{\Omega,\Omega}J)_{P,T}=0.\label{crit3}
\end{eqnarray}
To obtain the critical point one can use the first condition, which has been adopted in many works. In fact, the other two are also effective for obtaining the critical point. Next, we will show that the exact value of the critical point can be well determined by the second condition.

Applying the same technique used in Ref. \cite{Wei3}, the generalized Clapeyron equations can be obtained for the Kerr-AdS black hole in the extended phase space:
\begin{eqnarray}
 \bigg(\frac{dP}{dT}\bigg)_{J}&=&\frac{S_{B}-S_{A}}{V_{B}-V_{A}}=\frac{\Delta S}{\Delta V},\label{Clapeyron}\\
 \bigg(\frac{dJ}{dP}\bigg)_{T}&=&\frac{V_{B}-V_{A}}{\Omega_{B}-\Omega_{A}}=-\frac{\Delta V}{\Delta \Omega},\label{Clapeyron2}\\
 \bigg(\frac{dT}{dJ}\bigg)_{P}&=&\frac{\Omega_{B}-\Omega_{A}}{S_{B}-S_{A}}=\frac{\Delta \Omega}{\Delta S}.\label{Clapeyron3}
\end{eqnarray}
They hold for a singly spinning Kerr-AdS black hole with any dimension $d$.

\section{$d=4$-dimensional Kerr-AdS black holes}
\label{KerrAdS}

In this section, we will reexamine the phase transition of the four-dimensional Kerr-AdS black hole. At firstly, we give a brief review of its thermodynamics.

The line element of the Kerr-AdS black hole is given by
\begin{eqnarray}
 ds^{2}=-\frac{\Delta}{\rho^{2}}\bigg(dt-\frac{a\sin^{2}\theta}{\Xi}d\varphi\bigg)^{2}
        +\frac{\rho^{2}}{\Delta}dr^{2}+\frac{\rho^{2}}{1-a^{2}/l^{2}\cos^{2}\theta}d\theta^{2}\nonumber\\
        +\frac{(1-a^{2}/l^{2}\cos^{2}\theta)\sin^{2}\theta}{\rho^{2}}\bigg(adt-\frac{r^{2}+a^{2}}{\Xi}d\varphi\bigg)^{2},
\end{eqnarray}
where the metric functions read
\begin{eqnarray}
 \rho^{2}&=&r^{2}+a^{2}\cos^{2}\theta,\quad
 \Xi=1-\frac{a^{2}}{l^{2}},\\
 \Delta&=&(r^{2}+a^{2})(1+r^{2}/l^{2})-2mr.
\end{eqnarray}
The thermodynamic quantities are
\begin{eqnarray}
 T&=&\frac{r_{h}}{4\pi(r_{h}^{2}+a^{2})}
   \left(1+\frac{a^{2}}{l^{2}}+3\frac{r_{h}^{2}}{l^{2}}-\frac{a^{2}}{r_{h}^{2}}\right),\\
 S&=&\frac{\pi (r_{h}^{2}+a^{2})}{\Xi},\quad \Omega=\frac{a\Xi}{r_{h}^{2}+a^{2}}+\frac{a}{l^{2}},
\end{eqnarray}
where $r_{h}$ is the horizon radius of the black hole. The mass $M$ and angular momentum $J$ are related to the parameters $m$ and $a$ as
\begin{eqnarray}
 M=\frac{m}{\Xi^{2}},\quad J=\frac{am}{\Xi^{2}}.
\end{eqnarray}
The Gibbs free energy of the black hole is
\begin{eqnarray}
 G=\frac{a^4 \left(r_{h}^2-l^{2}\right)+a^2\left(3l^{4}
   +2l^{2}r_{h}^2+3r_{h}^4\right)+l^{2} r_{h}^2
   \left(l^{2}-r_{h}^2\right)}{4 r_{h}\left(a^2-l^{2}\right)^2}.
\end{eqnarray}
Moreover, we can express the temperature, Gibbs free energy, and thermodynamic volume in terms of $S$, $J$, and $P$:
\begin{eqnarray}
 T&=&\frac{S^2 \left(64 P^2S^2+32PS+3\right)
     -12\pi^2 J^2}{4\sqrt{\pi} S^{3/2} \sqrt{8 P S+3} \sqrt{12 \pi^2 J^2+S^2(8 P S+3)}},\label{TT}\\
 G&=&\frac{12\pi^2 J^2(16PS+9)-64 P^2 S^4+9S^2}{12\sqrt{\pi}\sqrt{S}\sqrt{8PS+3}
    \sqrt{12\pi^2J^2+S^2 (8PS+3)}},\\
 V&=&\frac{4\sqrt{S}\left(6\pi^2J^2+S^2(8PS+3)\right)}{3\sqrt{\pi}\sqrt{8PS+3}
   \sqrt{12\pi^2J^2+S^2(8PS+3)}}.\label{VV}
\end{eqnarray}
The specific volume can be determined with $v=2(3V/4\pi)^{1/3}$.

In general, the heat capacity implies a local thermodynamic stability: positive and negative heat capacities correspond to the stable and unstable systems, respectively. It should also be noted that the phase transition takes place at the divergent point of the heat capacity. For a rotating Kerr-AdS black hole, the heat capacity at fixed $J$ and $P$ is
\begin{eqnarray}
 C_{J,P}&=&T\bigg(\frac{\partial S}{\partial T}\bigg)_{J,P}\\
        &=&\frac{8T\sqrt{\pi}S^{5/2}\left((8PS+3)
   (12\pi^2J^2+S^2(8PS+3))\right)^{3/2}}
   {144\pi^4J^4(32PS+9)+24\pi^2J^2S^2(8PS+3)^2 (16 P
   S+3)+S^4(8PS-1)(8PS+3)^3}.\nonumber\label{heat}
\end{eqnarray}

\subsection{Analytical critical point}

It is shown that a small/large black hole phase transition of the vdW type exists in the background of the charged or rotating AdS black hole. For the charged AdS black hole, the thermodynamic volume depends only on the black hole horizon radius. Thus, the critical point can be easily determined by the condition (\ref{crit1}). However, for the rotating AdS black hole, the thermodynamic volume depends both on the angular momentum and the radius, which increases the difficulty of obtaining the critical point with Eq. (\ref{crit1}).

In Ref. \cite{Gunasekaran}, the authors expanded all quantities to $\mathcal{O}(J^{2})$ in the small-$J$ limit and obtained an approximate critical point, which reads
\begin{eqnarray}
 P_{c}^{a}=\frac{1}{12\sqrt{90}\pi}\frac{1}{J},\quad
 v_{c}^{a}=2\times 90^{1/4}\sqrt{J}, \quad
 T_{c}^{a}=\frac{90^{3/4}}{225\pi}\frac{1}{\sqrt{J}}.
\end{eqnarray}
The relation in turn yields the universal critical ratio $\rho_1=5/12$, which is different from the value $3/8$ for the vdW fluid. However, the critical exponents are the same as the vdW fluid, i.e., $\alpha=0$, $\beta=1/2$, $\gamma=1$, $\delta=3$. In Ref. \cite{Altamirano3}, the authors further approximated the exact equation of state by the slow-rotating expansion to a higher order. The critical point is slightly shifted from the preceding values, and the universal critical ratio $\rho_{1}$ deviates from $5/12$ by a small amount. However, the result shows that the critical exponents are the same as the ones obtained before.

Although Refs. \cite{Gunasekaran,Altamirano3} provided a good approximation to the exact equation of state, and reproduced the critical point, these results are based on the small-$J$ limit. One may wish to obtain the exact value of the critical point for any $J$. In the following, we will reexamine this issue.

Recall that the reason we could not obtain the exact critical point is mainly due to the fact that the thermodynamic and specific volumes are $J$ dependent (see Eqs. (\ref{TT}) and (\ref{VV})). This will make the condition (\ref{crit1}) hard to deal with, and the exact critical point cannot be obtained. However, as we shown before, there also exist other conditions to determine the critical point (see Eqs. (\ref{crit2}) and (\ref{crit3})).

In Ref. \cite{Tsai}, the authors studied the $J$-$\Omega$ criticality for the $d=4$-dimensional Kerr-AdS black hole. The condition (\ref{crit3}) was used, but analytical critical points were still not obtained. Thus the condition (\ref{crit2}) may be an alternative choice. Combined with Eq. (\ref{TT}), the condition $(\partial_{S}T)_{J,P}=(\partial_{S,S}T)_{J,P}=0$ reduces to
\begin{eqnarray}
 &&144\pi^4J^4(32PS+9)+24\pi^2J^2S^2(8PS+3)^2(16PS+3)+S^4(8PS-1)(8PS+3)^3=0,\\
 &&5184\pi^6J^6\left(512P^2S^2+288PS+45\right)-144\pi^4J^4S^2
   \left(32768P^4S^4-20160P^2S^2-8640PS-1053\right)\nonumber\\
  &&\quad+60\pi^2J^2S^4(8PS+3)^4(32PS+9)+S^6(8PS-3)(8PS+3)^5=0.
\end{eqnarray}
Recalling that the critical point must be in the form of Eq. (\ref{criticalpoint}), we only need to determine the dimensionless constants $\beta_{1}\sim\beta_{5}$. Fortunately, the above two complicated equations have an analytical solution:
\begin{eqnarray}
 P_{c} &=&\frac{k_{1}}{\pi}\cdot\frac{1}{J}\simeq\frac{0.002857}{J},\\
 v_{c} &=&2\bigg(\frac{\sqrt{k_{2}}\left(8k_{1}k_{2}^3+3k_{2}^2+6\right)}
       {\sqrt{8k_{1}k_{2}+3}\sqrt{8k_{1}k_{2}^3+3k_{2}^2+12}}\bigg)^{1/3}\cdot\sqrt{J}\simeq6.047357\sqrt{J},\\
 T_{c} &=&\frac{64k_{1}^2k_{2}^4+32k_{1}k_{2}^3+3k_{2}^2-12}
        {4\pi k_{2}\sqrt{k_{2}(8k_{1}k_{2}+3) \left(8k_{1}k_{2}^3+3k_{2}^2+12\right)}}\cdot\frac{1}{\sqrt{J}}\simeq\frac{0.041749}{\sqrt{J}},\\
 V_{c} &=&\frac{\pi}{6}v_{c}^{3}\simeq 115.796503 \sqrt[3]{J},
\end{eqnarray}
where
\begin{eqnarray}
 k_{1}&=&\frac{1}{64\left(103-3\sqrt{87}\right)^{17/3}}\times\bigg(-2^{2/3}\left(225679003807-24183767608\sqrt{87}\right)
   \sqrt[3]{103-3 \sqrt{87}}\nonumber\\
    &-&17\left(103-3\sqrt{87}\right)^{2/3}
   \left(484826973\sqrt{87}-5116133497\right)-\sqrt[3]{2}\left(68098470527+5855463275
   \sqrt{87}\right)\bigg),\nonumber\\
  k_{2}&=&\frac{2}{3}\left(2+\sqrt[3]{206-6\sqrt{87}}+\sqrt[3]{206+6\sqrt{87}}\right).\nonumber
\end{eqnarray}
Through examining the divergent behavior of the heat capacity $C_{J, P}$, Dolan \cite{DolanDolan} found the exact numerical value of the critical point for the $d=4$-dimensional Kerr-AdS black hole, which is consistent with our analytical result. It is also worth noting that the derivation of the analytical critical point has no approximation, and thus the result is valid for any value of $J$. In order to make a comparison with the approximate one given in Ref. \cite{Gunasekaran} (in the small-$J$ limit), we define the relative deviation $\Delta A_{c}=(A_{c}-A_{c}^{a})/A_{c}$ for a thermodynamic quantity $A_{c}$, with $A_{c}^{a}$ denoting the critical point obtained in the small-$J$ limit. Then we get
\begin{eqnarray}
 \Delta v_{c}=-1.87\%,\quad \Delta T_{c}=0.98\%,\quad
 \Delta P_{c}=2.13\%,\quad \Delta \rho_{1}=-0.69\%,
\end{eqnarray}
from which we can see that the deviations between the analytical and approximate values of the critical point are very small, i.e., within the range of 2.2\%. The analytical values of $v_{c}$ and $\rho_{1c}$ are smaller than the approximate ones, while $T_{c}$ and $P_{c}$ are larger. A special property of these deviations is that they are independent of the angular momentum $J$. This property seems very difficult to understand because that the approximate value of the critical point is obtained in the small-$J$ limit, while the analytical ones are effective for all $J$. The reason is that the Kerr-AdS black hole is a single-characteristic-parameter system, and its critical point has a unique form. With a simple calculation, one can confirm that the critical exponents and the scaling laws near such an exact critical point are the same as the ones obtained before.

\subsection{Coexistence curve and phase diagram}

There are classical swallowtail behavior of Gibbs free energy and $P$-$V$ oscillatory behavior when the phase transition takes place. In this subsection, we will study the phase transition information encoded in the thermodynamic quantities.

\begin{figure}
\center{\subfigure[]{\label{gtsa}
\includegraphics[width=6cm,height=4.8cm]{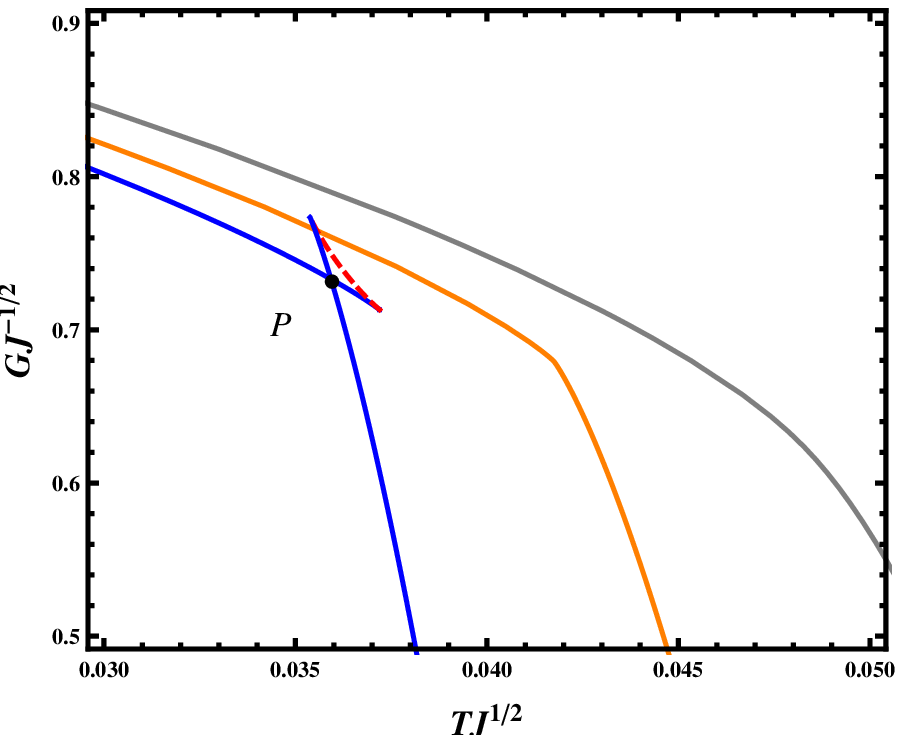}}
\subfigure[]{\label{}
\includegraphics[width=6cm,height=5cm]{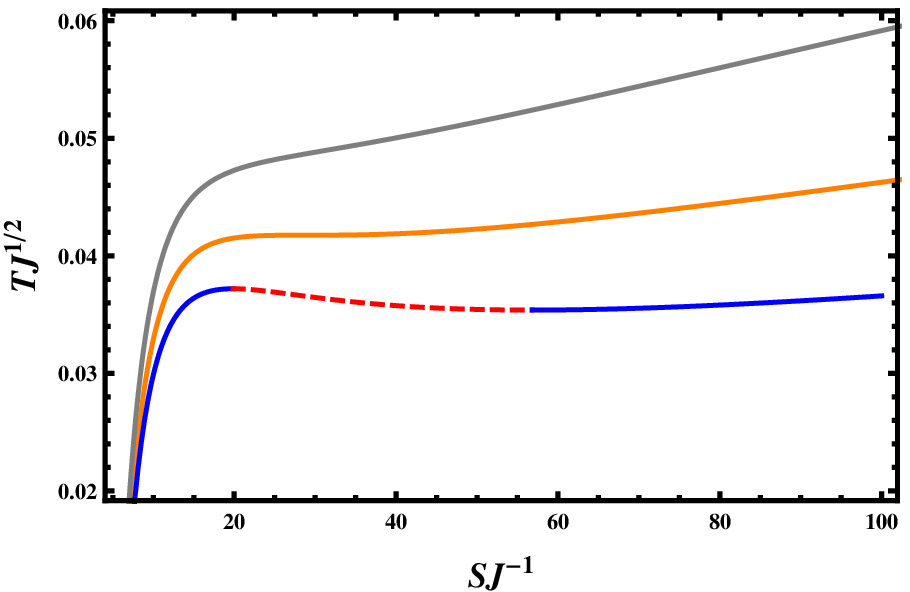}}}
\caption{(a) $G$ vs $T$. (b) $T$ vs $S$. The values of the pressure are set as $P=0.002$, $P_{c}$, and 0.004 from bottom to top. The dashed lines correspond to the black hole branch with negative heat capacity.}\label{gts}
\end{figure}

In fact, as shown in Refs. \cite{Spallucci,Smailagic}, the phase transition information is also encoded in the $T$-$S$ oscillatory behavior. We show the swallowtail behavior of $G$ and $the T$-$S$ oscillatory behavior for different values of the pressure $P$ in Fig. \ref{gts}. We conclude that the existence of the swallowtail behavior implies the $T$-$S$ oscillatory behavior. And the deflection point of $G$ corresponds to the extremal point of the $T$-$S$ oscillatory behavior. However, when the pressure $P>P_{c}$, both the swallowtail behavior and $T$-$S$ oscillatory behavior disappear.

For small pressure $P$, there are two extremal points on one $T$-$S$ curve, and the two points coincide with each other when $P=P_{c}$. When $P>P_c$, there is no extremal point. The extremal points are determined by
\begin{eqnarray}
 \bigg(\frac{\partial T}{\partial S}\bigg)_{J,P}=0.\label{extremal}
\end{eqnarray}
Thus the critical point occurs at the multiple root of the above equation. For a black hole with $T>0$, we have
\begin{eqnarray}
 C_{J,P}&=&T\bigg(\frac{\partial S}{\partial T}\bigg)_{J,P}
       \propto \bigg(\frac{\partial S}{\partial T}\bigg)_{J,P}.
\end{eqnarray}
Thus, it is clear that the heat capacity $C_{J,P}$ diverges at the extremal points determined by Eq. (\ref{extremal}). Moreover, it is known that a negative slope of the $T$-$S$ curve corresponds to a negative $C_{J,P}$, which implies unstable thermodynamics. Since the intermediate black hole branch has a negative slope on the $T$-$S$ curve, it should be excluded from the thermodynamic viewpoint.

%%%%%%%%%%%%%%%%%%%%%%%%%%%%%%%%%%%%%%%%%%%%%%%%%%%
%%%%%%%%%%%%%%%%%%%%%%%%%%%%%%%%%%%%%%%%%%%%%%%%%%%%%%%%%%%%%%%%%%%%%%%%%

The ability to exclude such a branch comes from the consideration that a thermodynamic system always prefers a lower Gibbs free energy. When $P<P_{c}$ (see Fig. \ref{gtsa}), as $T$ increase the system first prefers a small black hole branch until the point ``P" is reached, where the small and large black hole branches have the same value of $G$. So it is a coexistence point of the small and large black holes. As $T$ increase further, the large black hole branch will have a lower Gibbs free energy, and thus the system will prefer this branch. Since the intermediate black hole branch has a higher Gibbs free energy, it will be naturally excluded.

\begin{figure}
\center{\subfigure[]{\label{Coextence2a}
\includegraphics[width=7cm,height=6cm]{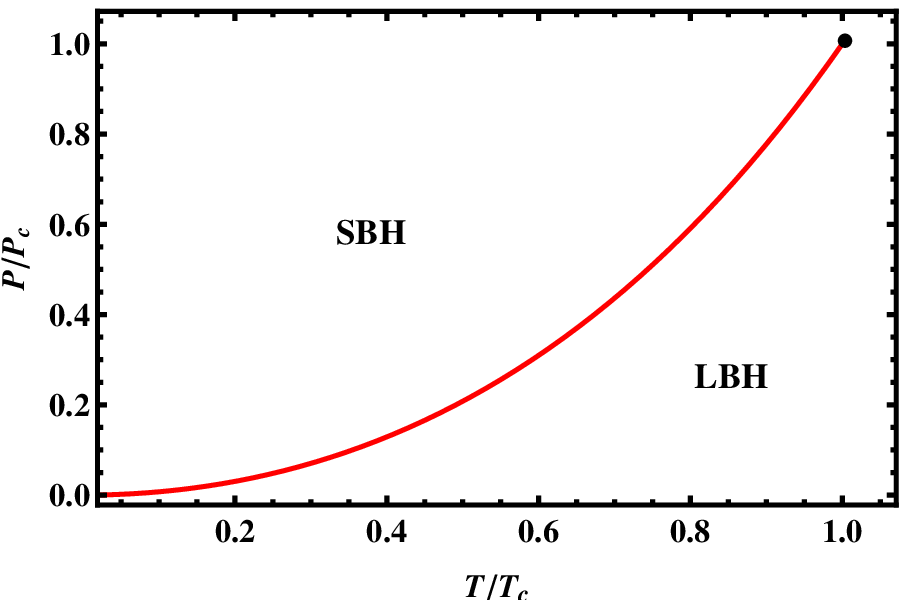}}
\subfigure[]{\label{PV2b}
\includegraphics[width=7cm,height=6cm]{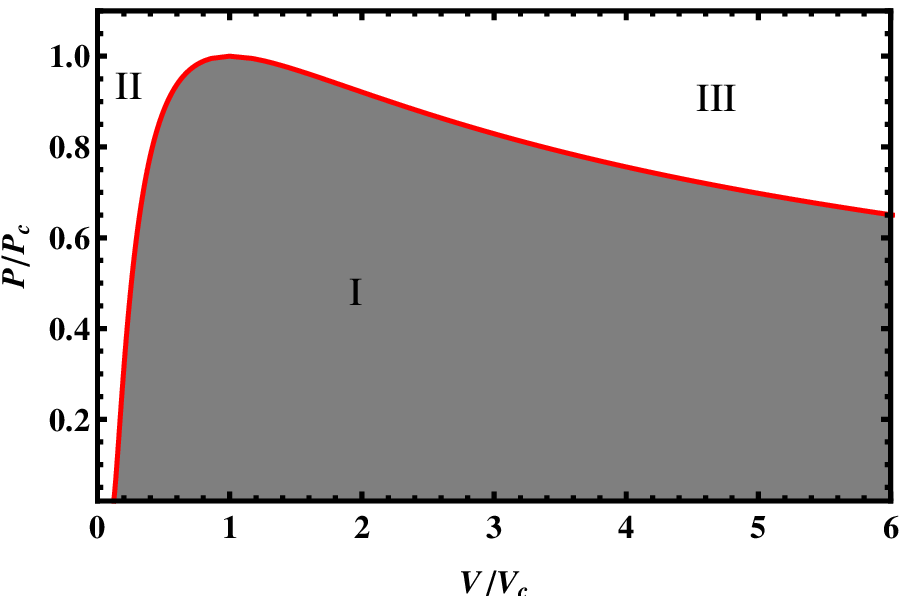}}}
\caption{Phase diagram in the reduced parameter space. (a) $P$-$T$ phase diagram. The red line denotes the coexistence curve. The region above it is the small black hole (SBH) phase, and the region below it is the large black hole (LBH) phase. The dot located at (1, 1) corresponds to the critical point. (b) $P$-$V$ phase diagram. Region I is the coexistence phase of the small and large black holes. Regions II and III are the small and large black hole phases, respectively.}
\end{figure}

In this process, ``P" is a special point, at which the small and large black holes can coexist. This point can also be equivalently determined by the equal-area laws given above. By letting the parameters $J$ and $P$ vary independently, we will obtain all the coexistence points. For the charged AdS black hole, it was shown in Ref. \cite{Wei3} that the coexistence curve is charge independent in the reduced parameter space for any dimension $d$. For the rotating Kerr-AdS black hole, we wonder whether the coexistence curve is angular momentum $J$ independent. With this issue, we carry out the same calculation. For the rotating Kerr-AdS black hole, the parametrization form of the coexistence curve reads
\begin{eqnarray}
 \tilde{P}&=&0.718781
   \tilde{T}^2+0.188586 \tilde{T}^3+0.061488 \tilde{T}^4+0.022704
   \tilde{T}^5\nonumber\\
   &+&0.002340 \tilde{T}^6+0.010547 \tilde{T}^7-0.008649 \tilde{T}^8+0.005919
   \tilde{T}^9-0.001717 \tilde{T}^{10},  \label{coex4}
\end{eqnarray}
where the reduced parameters are defined as
\begin{eqnarray}
 \tilde{P}=\frac{P}{P_{c}},\quad \tilde{T}=\frac{T}{T_{c}}.
\end{eqnarray}
It worth noting that for the rotating Kerr-AdS black hole, there is a $J$-independent behavior for the coexistence curve in the reduced parameter space. To be clear, we plot the coexistence curve in the reduced parameter space in Fig. \ref{Coextence2a}. This coexistence curve has a positive slope everywhere and terminates at the critical point. Above this curve is the range of the small black hole, and below it is that of the large black hole. On the curve, the small and large black holes can coexist. With the parametrization form of the coexistence curve (\ref{coex4}), we can check that the Clapeyron equations (\ref{Clapeyron})-(\ref{Clapeyron3}) hold. The $P$-$V$ phase diagram is presented in Fig. \ref{PV2b}.

%%%%%%%%%%%%%%%%%%%%%%%%%%%%%%%%%%%%%%%%%%%%%%%%%%%%%%%%%%%%%%%%%%%%%
%%%%%%%%%%%%%%%%%%%%%%%%%%%%%%%%%%%%%%%%%%%%%%%%%%%%%%%%%%%%%%%%%%%%%%%%%

%%%%%%%%%%%%%%%%%%%%%%%%%%%%%%%%%%%%%%%%%%%%%%%%%%%%%%%%%%%%%%%%%%%%%
\begin{figure}
\center{\subfigure[]{\label{TVphasea}
\includegraphics[width=7cm,height=6cm]{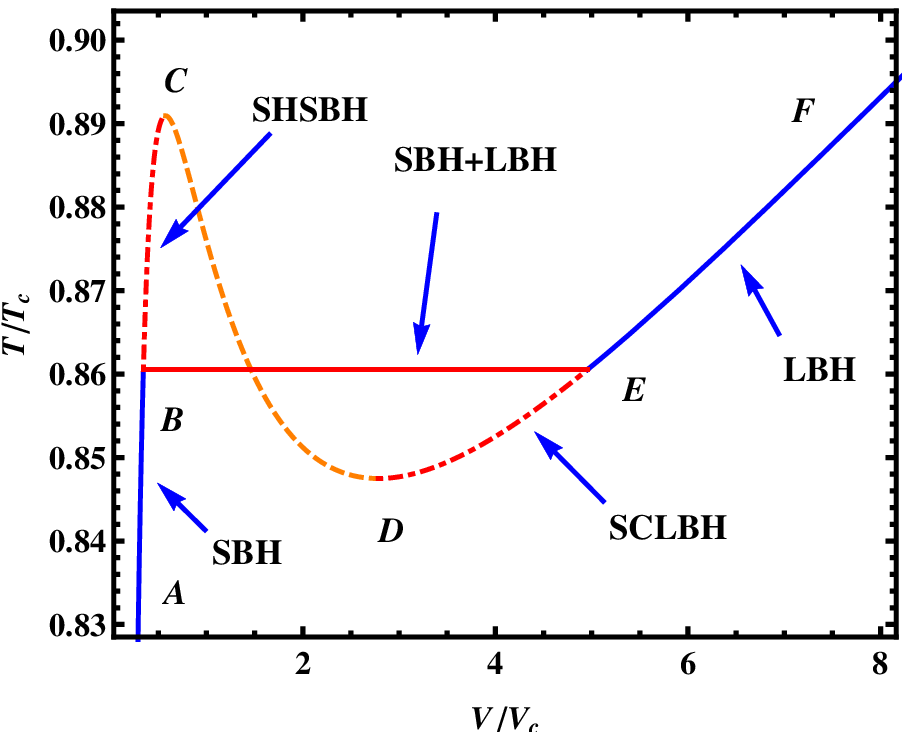}}
\subfigure[]{\label{TVphaseb}
\includegraphics[width=7cm,height=6cm]{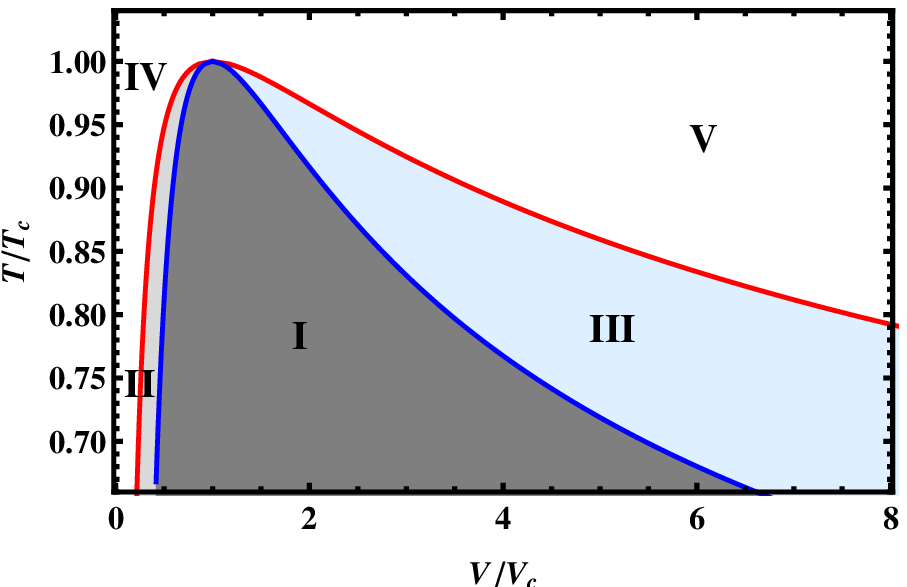}}}
\caption{(a) Isobaric line in the $T$-$V$ plane with $P/P_{c}$=0.7. The phase transition temperature for this case is of about $T/T_{c}=0.86$. Branches AC, CD, DF, respectively, correspond to small, intermediate, and large black hole phases. Branches BC and DE denote the superheated small black hole (SHSBH) and supercooled large black hole (SCLBH) phases. (b) $T$-$V$ phase diagram. The region in the shadow (I+II+III) denotes the coexistence phase of the small and large black holes. In this shadow region, there are two subphases, i.e., the SHSBH phase (II) and the SCLBH phase (III). Regions IV and V correspond to the small and large black hole phases.}
\end{figure}
%%%%%%%%%%%%%%%%%%%%%%%%%%%%%%%%%%%%%%%%%%%%%%%%%%%%%%%%%%%%%%%%%%%%%%%%%

Another alternative way to study the phase transition is the $T$-$V$ phase diagram. In Fig. \ref{TVphasea}, we plot the isobaric line in the reduced parameter space with $P/P_{c}$=0.7. There is also an oscillatory behavior. The red horizontal line denotes the temperature of the phase transition calculated from the Gibbs free energy, where the small and large black holes coexist. These lines construct two areas. However, with a simple calculation, we find that these two areas are not equal. So there is no equal-area law on the $T$-$V$ oscillatory line.

The branch ``CD" with a negative slope corresponds to the intermediate black hole. We remove this branch by replacing the ``BE" oscillatory line with the red horizontal line. Then the black hole system follows the path ``A-B-E-F" as the size of the black hole increase. When the size decreases, the system follows the path ``F-E-B-A". In fact, there is another path for the system to follow. It first starts from point ``A" and moves to point ``B". Then, due to the density fluctuations or interactions, the system will not proceed to ``E", but rather to ``C". In analogy to vdW fluids, we can call the system located at the ``BC" branch the superheated small black hole. Then as the temperature increases further, this small black hole system will rapidly change to a large black hole system, resulting in an explosion. If the system starts at point ``F" and gets to point ``E" by decreasing its size, it has a chance to arrive at point ``D". This large black hole branch ``ED" corresponds to a low temperature, and we can call it a supercooled large black hole as it has lower temperature than the other large black hole.

We show the $T$-$V$ phase diagram in the reduced parameter space in Fig. \ref{TVphaseb}. One can see that there are five different black hole phases. Region ``I" (gray) in gray color is a coexistence phase, where large and small black holes coexist. Regions ``IV" and ``V" correspond to small and large black hole phases, respectively. Regions ``II" (light gray) and ``III" (light blue) are two subphases, i.e., superheated small black hole phase and supercooled large black hole phase, which do not appear in the $P$-$T$ phase diagram.

\section{$d\geq 5$-dimensional Kerr-AdS black holes}

Here we would like to consider a high-dimensional Kerr-AdS black hole with only one nonvanishing rotation parameter. The line element of the $d$-dimensional singly spinning Kerr-AdS black hole is
\begin{eqnarray}
 ds^{2}=&-&\frac{\Delta}{\rho^{2}}\bigg(dt-\frac{a\sin^{2}\theta}{\Xi}d\varphi\bigg)^{2}
        +\frac{\rho^{2}}{\Delta}dr^{2}+\frac{\rho^{2}}{1-a^{2}/l^{2}\cos^{2}\theta}d\theta^{2}\nonumber\\
        &+&\frac{(1-a^{2}/l^{2}\cos^{2}\theta)\sin^{2}\theta}{\rho^{2}}\bigg(adt-\frac{r^{2}+a^{2}}{\Xi}d\varphi\bigg)^{2}
        +r^{2}\cos^{2}\theta d\Omega_{d-4}^{2},
\end{eqnarray}
where $d\Omega_{d-4}$ denotes the metric element on a $d$-dimensional
sphere and
\begin{eqnarray}
 \Delta=(r^{2}+a^{2})(1+\frac{r^{2}}{l^{2}})-2mr^{5-d},\\
 \Xi=1-\frac{a^{2}}{l^{2}},\quad \rho^{2}=r^{2}+a^{2}\cos^{2}\theta.
\end{eqnarray}
The thermodynamic quantities are
\begin{eqnarray}
 M&=&\frac{\omega_{d-2}}{4\pi}\frac{m}{\Xi^{2}}\bigg(1+\frac{(d-4)}{2}\Xi\bigg),\quad
 J=\frac{\omega_{d-2}}{4\pi}\frac{ma}{\Xi^{2}},\\
 \Omega&=&\frac{a}{l^{2}}\frac{r_{h}^{2}+l^{2}}{r_{h}^{2}+a^{2}},\quad
 S=\frac{A}{4}=\frac{\omega_{d-2}}{4}\frac{(a^{2}+r_{h}^{2})r_{h}^{d-4}}{\Xi},\label{th}\\
 T&=&\frac{r_{h}}{2\pi}\bigg(\frac{r_{h}^{2}}{l^{2}}+1\bigg)\bigg(\frac{1}{r_{h}^{2}+a^{2}}+\frac{d-3}{2r_{h}^{2}}\bigg)
 -\frac{1}{2\pi r_{h}},\\
 G&=&\frac{\omega_{d-2}r_{h}^{d-5}}{16\pi \Xi^{2}}\bigg(3a^{2}+r_{h}^{2}-\frac{(r_{h}^{2}-a^{2})^{2}}{l^{2}}
  +\frac{3a^{2}r_{h}^{4}+a^{4}r_{h}^{2}}{l^{4}}\bigg),\\
 V&=&\frac{r_{h}A}{d-1}\bigg(1+\frac{a^{2}}{\Xi}\frac{1+r_{h}^{2}/l^{2}}{(d-2)r_{h}^{2}}\bigg),\quad
 v=\frac{V}{S},
\end{eqnarray}
where $\omega_{d}=2\pi^{(d+1)/2}/\Gamma((d+1)/2)$, and the horizon radius $r_{h}$ of the black hole is determined by $\Delta=0$. In the small-$J$ limit, the critical point has been obtained in Ref. \cite{Altamirano3} for any dimension $d$, which reads
\begin{eqnarray}
 T_{c}=\frac{4(d-3)}{\pi(2d-3)}\frac{1}{b},\quad
 P_{c}=\frac{d-3}{\pi(d-1)}\frac{1}{b^{2}},\quad
 b=\frac{4}{d-2}\bigg(\frac{2^{6}\pi^{2}(2d-3)(d-1^{2}J^{2})}{(d-2)(d-3)\omega_{d-2}^{2}}\bigg)^{1/2(d-2)}.
 \label{appcrit}
\end{eqnarray}

As shown above, for the four-dimensional Kerr-AdS black hole, we have the analytical critical point, but when the dimension $d$ is larger than four, there is no the analytical expression. However the critical point must be in the form of Eq. (\ref{criticalpoint}), and $\beta_{1}$---$\beta_{4}$ can be numerically determined.

In order to get the exact critical point, we proceed as follows. We first express $a=a(r_{h}, P, S)$ from the $S$ equation (\ref{th}) with $J=1$. Then we solve the $J$ equation to obtain $r_{h}=r_{h}(P, S)$, which is in a complex form and we will not show it here. Inserting them into the $T$ equation, one gets $T=T(P, S)$. At last, using the conditions (\ref{crit2}), we obtain the constants $\beta_{1}$---$\beta_{4}$, and the exact critical point is obtained.

\subsection{$d=$5-dimensional case}

It was shown in Ref. \cite{Altamirano3} that, similar to the $d=4$ case, there is also a small/large black hole phase transition of vdW type for $d=5$. We here take the $d=5$ case as an example to obtain the critical point following the process we presented above.

\begin{figure}
\subfigure[]{\label{5ts}
\includegraphics[width=7cm,height=6cm]{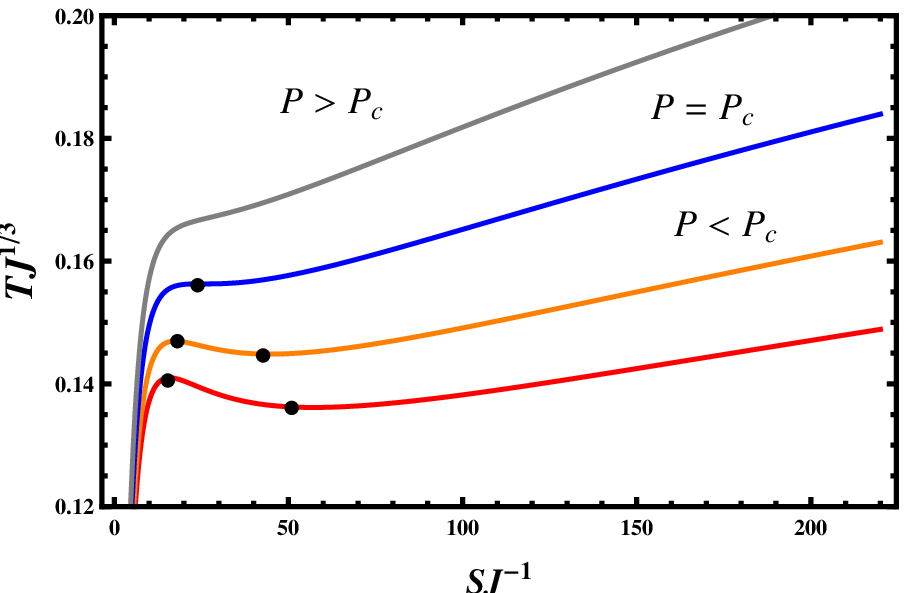}}
\subfigure[]{\label{5ps}
\includegraphics[width=7cm,height=6cm]{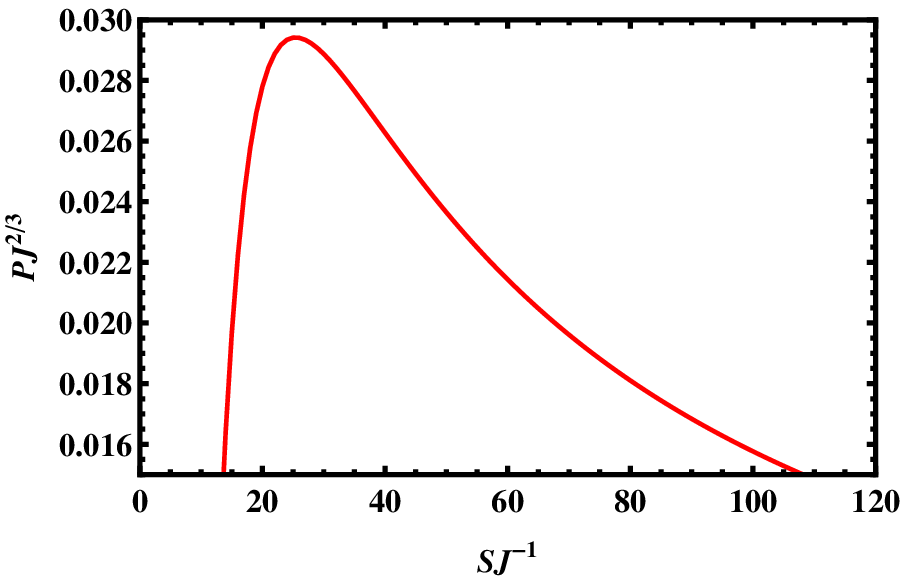}}
\caption{Five-dimensional Kerr-AdS black hole. (a) $T$ vs $S$ with fixed pressure $P$=0.022, 0.025, $P_{c}$, and 0.034 from bottom to top. Black dots denote the extremal points of the isobaric lines. When $P<P_{c}$, there are two extremal points, which coincide with each other at $P=P_{c}$, while when $P>P_{c}$, no extremal point exists. (b) The extremal point ($(\partial_{S}T)_{J,P}=0$) showed in the $P$-$S$ plane.}
\end{figure}

The state equation of $T=T(P, S)$ with $J=1$ is in a complex form, and thus we only show the figure for it. The isobar is shown in Fig. \ref{5ts}. It is clear that for $P<P_{c}$, there are $T$-$S$ oscillatory behavior and two extremal points. When $P=P_{c}$, the two extremal points coincide with each other corresponding to the critical case. With further increase of $P$, no extremal point exists. This behavior of the isobar is very similar to the isotherm of a vdW fluid. And we have pointed out in Ref. \cite{Wei3} that constructing the equal-area law also reproduces the exact phase transition. In order to obtain the critical point, we need to solve the two equations $(\partial_{S}T)_{J,P}=(\partial_{S,S}T)_{J,P}=0$. However, it is very difficult to directly solve them. Here we can solve the first equation $(\partial_{S}T)_{J,P}=0$, and then find the multiple root of it. Thus the critical point will be obtained. We plot the extremal point in Fig. \ref{5ps}. The result confirms the same situation as in Fig. \ref{5ts}, and when the pressure approaches about $0.029$, the two extremal points meet each other. Thus we can solve the second equation $(\partial_{S,S}T)_{J,P}=0$ near $P=0.029$. Finally, we obtain the exact critical point
\begin{eqnarray}
  P_{c}=0.029418J^{-2/3},\quad T_{c}=0.156284J^{-1/3}, \quad
  S_{c}=25.404144J,\quad V_{c}=43.874111J^{4/3}, \quad v_{c}=1.727045J^{1/3}.
\end{eqnarray}
The phase diagram is similar to the case of $d=4$, and we will not explore it.

%%%%%%%%%%%%%%%%%%%%%%%%%%%%%%%%%%%%%%%%%%%%%%%%%%%%%%%%%%%%%%%%%%%%%
%%%%%%%%%%%%%%%%%%%%%%%%%%%%%%%%%%%%%%%%%%%%%%%%%%%%%%%%%%%%%%%%%%%%%%%%%

\subsection{$d$=6-dimensional case}

For the $d$=6-dimensional singly spinning Kerr-AdS black hole, the reentrant phase transition emerges, which provides an opportunity to determine the phase structure of the AdS black hole. The study is also generalized to equal-spinning Kerr-AdS black holes in $d$-dimensions. However, the exact critical point is still missing. We would like to find the exact value for it with the method presented for the $d=5$ case.

%%%%%%%%%%%%%%%%%%%%%%%%%%%%%%%%%%%%%%%%%%%%%%%%%%%%%%%%%%%%%%%%%%%%%
\begin{figure}
\center{\includegraphics[width=8cm,height=6cm]{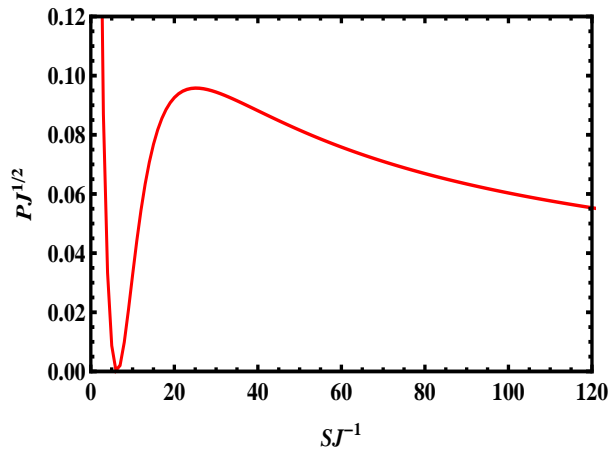}}
\caption{Behavior of the extremal point for the $d$=6-dimensional Kerr-AdS black hole.}\label{6ps}
\end{figure}
%%%%%%%%%%%%%%%%%%%%%%%%%%%%%%%%%%%%%%%%%%%%%%%%%%%%%%%%%%%%%%%%%%%%%%%%%

\begin{figure}
\subfigure[]{\label{6tsa}
\includegraphics[width=7cm,height=6cm]{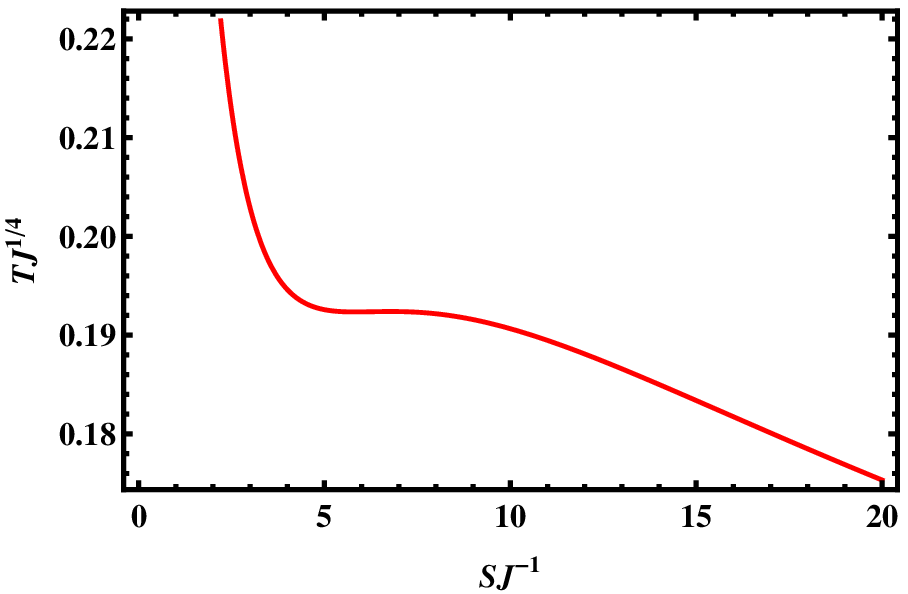}}
\subfigure[]{\label{6tsb}
\includegraphics[width=7cm,height=6cm]{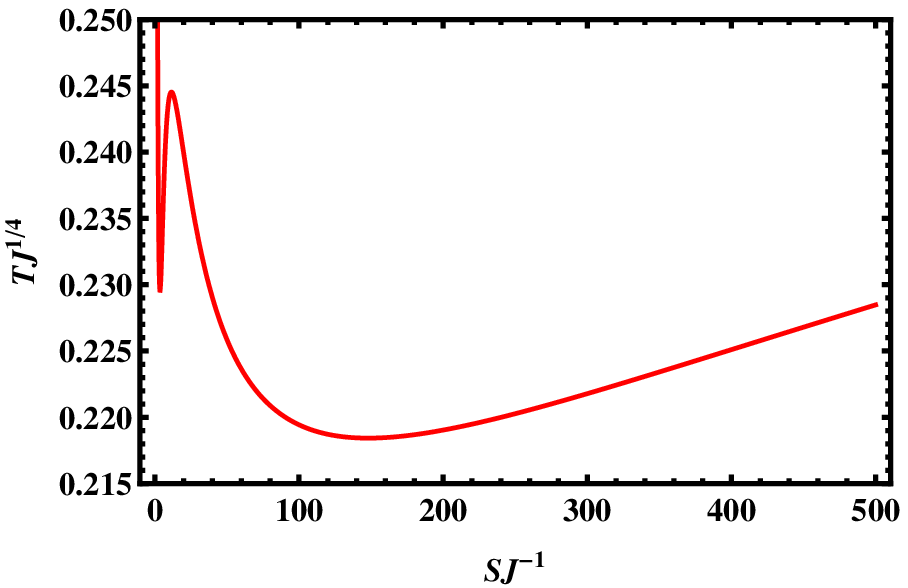}}\\
\subfigure[]{\label{6tsc}
\includegraphics[width=7cm,height=6cm]{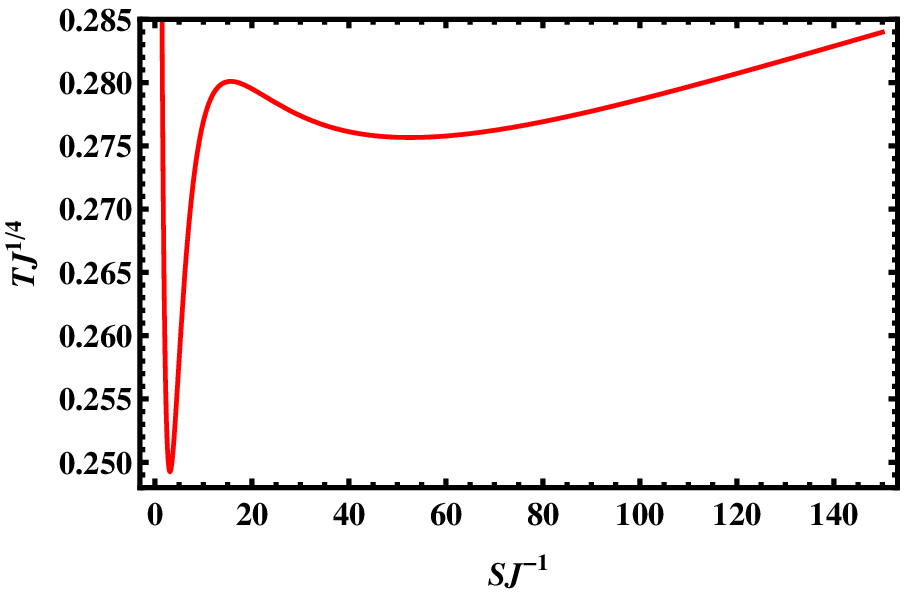}}
\subfigure[]{\label{6tsd}
\includegraphics[width=7cm,height=6cm]{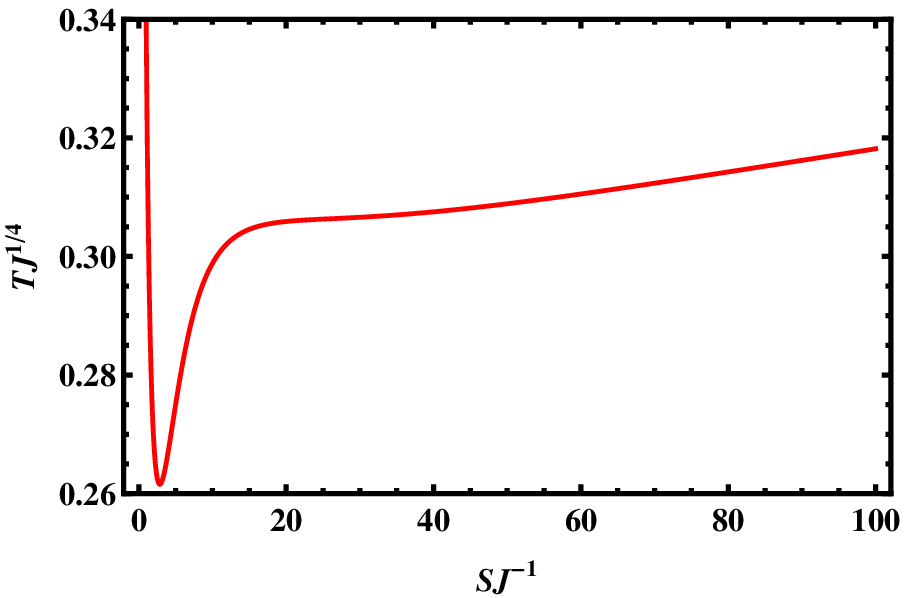}}
\caption{Different behaviors of $T$ as a function of $S$ with fixed $P$ for the $d$=6-dimensional singly spinning Kerr-AdS black hole. (a) $P$=0.001. (b) $P$=0.05, (c) $P$=0.08, and (d) $P$=0.1.}\label{p6ts}
\end{figure}

We fist list the extremal point in Fig. \ref{6ps}. From it, one can see that the structure of the extremal point is richer than the $d=5$ case implying a richer phase structure. This structure admits two critical points:
\begin{eqnarray}
  P^{(6)}_{c1}&=&0, \quad T^{(6)}_{c1}=,0.191463J^{-1/4},\nonumber\\
        S^{(6)}_{c1}&=&6.283185J, \quad V^{(6)}_{c1}=6.039680J^{5/4}, \quad v^{(6)}_{c1}=0.961245J^{1/4};\\
  P^{(6)}_{c2}&=&0.095793J^{-1/2}, \quad T^{(6)}_{c2}=0.300423J^{-1/4}, \nonumber\\
        S^{(6)}_{c2}&=&25.204559J,\quad V^{(6)}_{c2}=40.571880J^{5/4},\quad v^{(6)}_{c2}=1.609704J^{1/4}.
\end{eqnarray}
To be clear, we show the $T$-$S$ oscillatory behavior in Fig. \ref{p6ts}. The isobar behaves differently at different pressures (for details, one can see Refs. \cite{Altamirano3,Altamirano,AltamiranoKubiznak}). Considering that the black hole near the first critical point always has a higher Gibbs free energy than the very large black hole, the first critical point will not emerge in the phase diagram. However, if the very large black hole branch is suppressed by some other physical mechanisms, then the first critical point will appear in the phase diagram.

%%%%%%%%%%%%%%%%%%%%%%%%%%%%%%%%%%%%%%%%%%%%%%%%%%%%%%%%%%%%%%%%%%%%%
%%%%%%%%%%%%%%%%%%%%%%%%%%%%%%%%%%%%%%%%%%%%%%%%%%%%%%%%%%%%%%%%%%%%%%%%%

\subsection{$d\geq7$-dimensional case}

\begin{figure}
\center{\includegraphics[width=8cm,height=6cm]{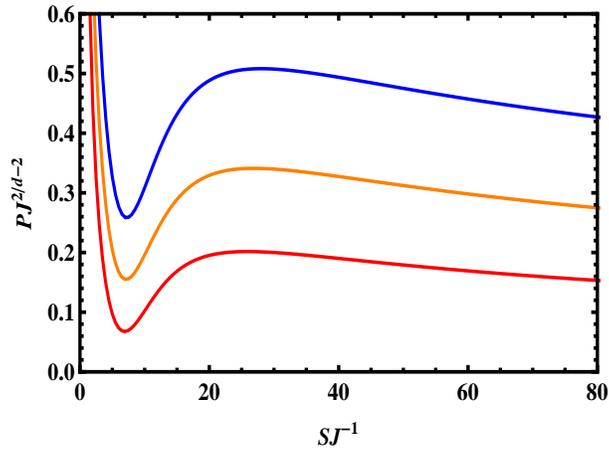}}
\caption{Extremal points of the Kerr-AdS black holes with $d$=7, 8, and 9 from bottom to top.}\label{789ps}
\end{figure}

$d\geq7$-dimensional Kerr-AdS black holes have similar critical phenomena as the $d=6$ case. Here we present the extremal points for $d=7, 8, 9$ in Fig. \ref{789ps}, which are very similar to the $d=6$ case shown in Fig. \ref{6ps}. There are two critical points for $d\geq7$, which we list in Table \ref{parameters}. Similar to the $d=6$ case, the first critical point does not participate in the phase transition. When $d\geq10$, the calculation becomes more difficult. However we believe that the situation is the same as $d=6$.

%%%%%%%%%%%%%%%%%%%%%%%%%%%%%%%%%%%%%%%%%%%%%%%%%%%%%%%%%%%%%%%%%%%%%
%%%%%%%%%%%%%%%%%%%%%%%%%%%%%%%%%%%%%%%%%%%%%%%%%%%%%%%%%%%%%%%%%%%%%%%%%

Before ending this section, we give a brief summary. For $d=5$---9, we obtain the exact critical points. When $d=5$, there is only one critical point, and the critical phenomena is the same as $d=4$. When $d\geq6$, there are two critical points, and the reentrant phase transition emerges. All the critical points are exact without any approximation. In Ref. \cite{Altamirano3}, the authors gave the approximate critical point (\ref{appcrit}) in the small-$J$ limit, which is consistent with the second critical point obtained by us. We list the relative deviations of the critical pressure and temperature in Table \ref{II}. It is clear that, the approximate one has small deviations $|\Delta P_{c}|$ and $|\Delta T_{c}|$ from the exact one. With an increase of the dimension $d$, the deviations decrease. Although the approximate critical points were obtained in the small-$J$ limit, they are still effective for large $J$, which is mainly because the singly spinning Kerr-AdS black hole system is a single-characteristic-parameter system.

%%%%%%%%%%%%%%%%%%%%%%%%%%%%%%%%%%%%%%%%%%%%%%%%%%%%%%%%%%%%%%%%%%%%%%%
\begin{table}[h]
\begin{center}
\begin{tabular}{c c c c c c }
  \hline
  \hline
   $d$&$\beta_{1}$ &$\beta_{2}$ & $\beta_{3}$ & $\beta_{4}$ & $\beta_{5}$  \\
  \hline
    7 & 0.067650 &0.334529 & 6.906010  & 4.363333 & 0.631817  \\
      & 0.201610 &0.451135 & 25.883786 & 30.196200 & 1.166607  \\\hline
    8 & 0.154941 &0.467142 & 7.135946 & 3.332207 & 0.466961 \\
      & 0.341077 &0.599656 & 26.887658 & 23.703602 & 0.881579  \\\hline
    9 & 0.258243&0.590912 & 7.225609 & 2.718716 & 0.376261  \\
      & 0.508084&0.742869 & 28.018164 & 19.905363 & 0.710445 \\
  \hline\hline
\end{tabular}
\end{center}
\caption{Coefficients of the critical points for $d=7$, 8, and 9-dimensional Kerr-AdS black holes. The form of the critical point is given in Eq.~(\ref{criticalpoint}).}\label{parameters}
\end{table}
%%%%%%%%%%%%%%%%%%%%%%%%%%%%%%%%%%%%%%%%%%%%%%%%%%%%%%%%%%%%%%%%%%%%%%%

%%%%%%%%%%%%%%%%%%%%%%%%%%%%%%%%%%%%%%%%%%%%%%%%%%%%%%%%%%%%%%%%%%%%%%%
\begin{table}[h]
\begin{center}
\begin{tabular}{c  c c c c c}
  \hline
  \hline
   $d$  &  5 & 6 & 7 & 8 & 9  \\
  \hline
    $|\Delta P_{c}|$(\%) &  2.34 & 1.95 & 1.55 & 1.23  & 0.99     \\
    $|\Delta T_{c}|$(\%) &  1.10 & 0.93 & 0.74 & 0.59  & 0.45      \\
  \hline\hline
\end{tabular}
\end{center}
\caption{The relative deviations of the critical pressure and temperature between the approximate one (\ref{appcrit}) and the exact one (second critical point given by us in Table \ref{parameters}).}\label{II}
\end{table}
%%%%%%%%%%%%%%%%%%%%%%%%%%%%%%%%%%%%%%%%%%%%%%%%%%%%%%%%%%%%%%%%%%%%%%%

\section{Exact critical reentrant phase transition point}
\label{reentrant}

\begin{figure}
\subfigure[]{\label{RGtza}
\includegraphics[width=7cm,height=6cm]{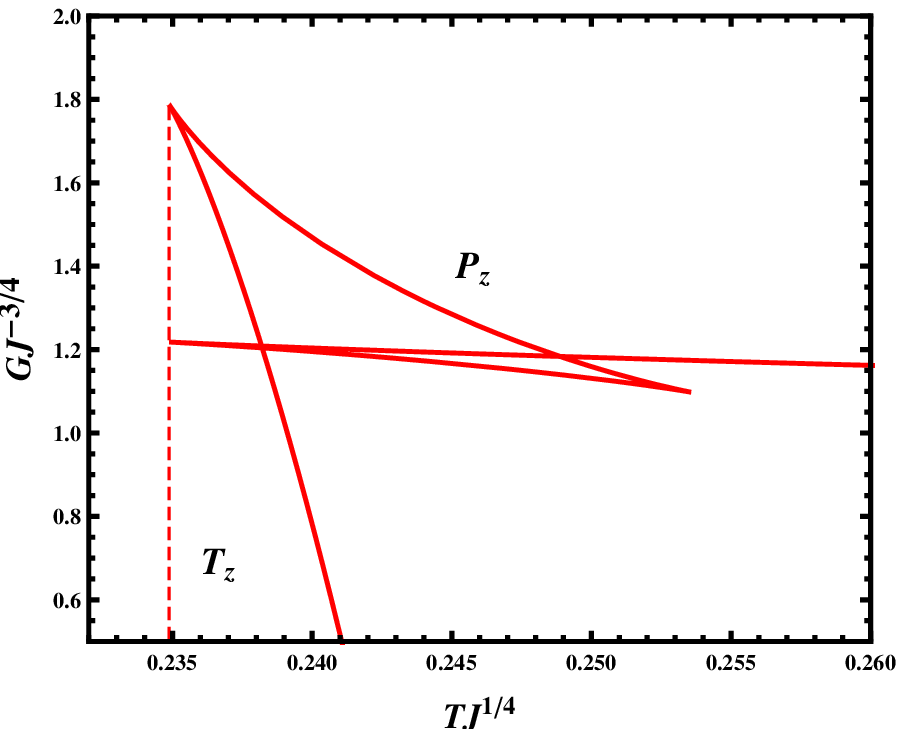}}
\subfigure[]{\label{RGtzb}
\includegraphics[width=7cm,height=6cm]{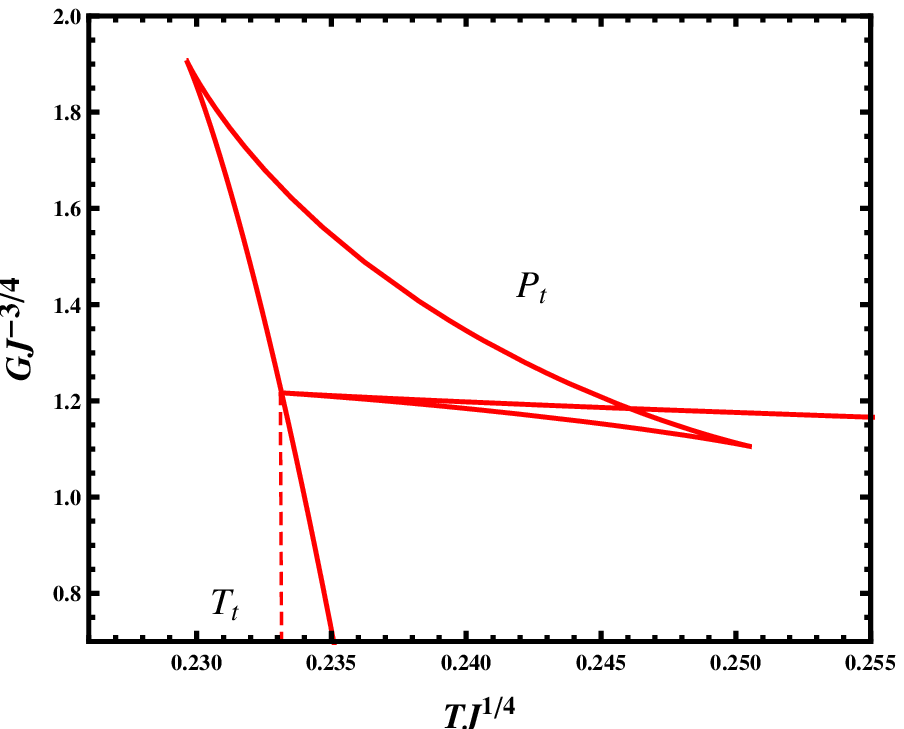}}
\caption{Characteristic behavior of the Gibbs free energy for the $d$=6 case. (a) $P=P_{z}$, (b) $P=P_{t}$.}\label{RGtz}
\end{figure}

It was clearly shown in Ref. \cite{Altamirano} that, when $d\geq6$, the reentrant phase transitions occur. The $P$-$T$ phase diagram can be found in Fig.~4 of Ref. \cite{Altamirano}. Decreasing the pressure from the second critical point, the reentrant phase transition occurs at $P=P_{z}$, and terminates at $P=P_{t}$. Here we call the points $P=P_{z}, P_{t}$ the CRPs. We plot the Gibbs free energy for $d$=6-dimensional singly spinning Kerr-AdS black hole at $P=P_{z}$ and $P=P_{t}$ in Fig. \ref{RGtz}, from which one can clearly see a multicharacteristic swallowtail behavior.

%%%%%%%%%%%%%%%%%%%%%%%%%%%%%%%%%%%%%%%%%%%%%%%%%%%%%%%%%%%%%%%%%%%%%
%%%%%%%%%%%%%%%%%%%%%%%%%%%%%%%%%%%%%%%%%%%%%%%%%%%%%%%%%%%%%%%%%%%%%%%%%

\begin{figure}
\center{\includegraphics[width=8cm,height=6cm]{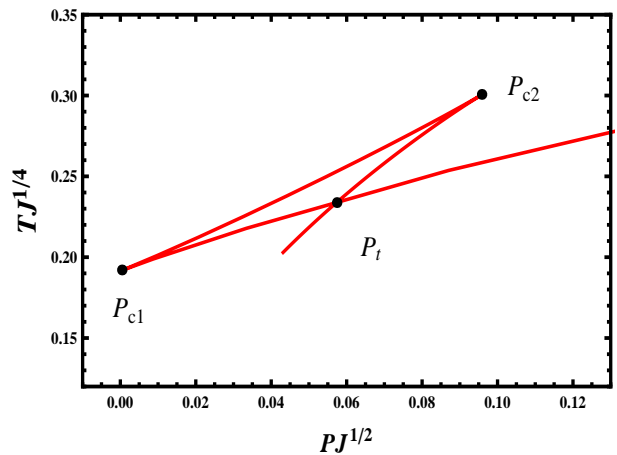}}
\caption{Extremal point of $(\partial_{S}T)_{J,P}$ in the $T$-$P$ plane for $d=6$. From this figure, we can read out the two critical points and the CRP ($P_{t}, T_{t}$).}\label{RT123}
\end{figure}

Using the dimensional analysis from Sec. \ref{Classification}, the CRPs have the following form:
\begin{eqnarray}
 P_{z}&=&\gamma_{1} J^{-\frac{2}{d-2}},\quad
 T_{z}=\gamma_{2}J^{-\frac{1}{d-2}},\nonumber\\
 P_{t}&=&\delta_{1} J^{-\frac{2}{d-2}},\quad
 T_{t}=\delta_{2}J^{-\frac{1}{d-2}}.\label{CRPs}
\end{eqnarray}
From the properties of $T$ and $G$, we find that at constant pressure the deflection point of $G$ is exactly related to the extremal point of $T$. Thus from Fig. \ref{RGtza}, we see that at the CRP ($P_{z}, T_{z}$), the two extremal points of $(\partial_{S}T)_{J,P}=0$ have the same value of $T$. According to this result, we can numerically determine the CRP ($P_{z}, T_{z}$). First, we show the extremal points of $T$ in the $T$-$P$ plane for $d=6$ in Fig. \ref{RT123}, and the cases for $d=7$---9 are the same. The two critical points are located at the deflection points. At the intersection point, the two extremal points share the same $T$ and $P$, which is just the CRP ($P_{z}, T_{z}$). By finding this point, we will get the value of the CRP. There is no good method to obtain the exact value of CRP ($P_{t}, T_{t}$), so we get the value only through fine-tuning by hand. In Table \ref{table3}, we list the exact values of the CRPs for $d=6$---9.

%%%%%%%%%%%%%%%%%%%%%%%%%%%%%%%%%%%%%%%%%%%%%%%%%%%%%%%%%%%%%%%%%%%%%
%%%%%%%%%%%%%%%%%%%%%%%%%%%%%%%%%%%%%%%%%%%%%%%%%%%%%%%%%%%%%%%%%%%%%%%%%

%%%%%%%%%%%%%%%%%%%%%%%%%%%%%%%%%%%%%%%%%%%%%%%%%%%%%%%%%%%%%%%%%%%%%%%
\begin{table}[h]
\begin{center}
\begin{tabular}{c  c c c c }
  \hline
  \hline
   $d$  &  $\gamma_{1}$ & $\gamma_{2}$ & $\delta_{1}$ & $\delta_{2}$  \\
  \hline
    6 &  0.057850 & 0.234874 & 0.055279 & 0.233149 \\
    7 &  0.146703 & 0.386293 & 0.142321 & 0.383623 \\
    8 &  0.263912 & 0.528891 & 0.257690 & 0.525563 \\
    9 &  0.404161 & 0.663894 & 0.396070 & 0.660045 \\
  \hline\hline
\end{tabular}
\end{center}
\caption{Exact values of the coefficients of the CRPs for $d=6$---9. The explicit forms are given in Eq.~(\ref{CRPs}).}\label{table3}
\end{table}
%%%%%%%%%%%%%%%%%%%%%%%%%%%%%%%%%%%%%%%%%%%%%%%%%%%%%%%%%%%%%%%%%%%%%%%

\section{Conclusions}
\label{Conclusion}

In this paper, we have studied the critical phenomena of the $d$-dimensional singly spinning Kerr-AdS black holes. We first examined the thermodynamic quantities for such a system, and they are divided into two classes. The first class consists of the universal parameters, such as $P$, $V$, and $T$. The second class consists of the characteristic parameters, such as the $Q$ and $J$. The phase transition and critical phenomenon are dependent on the characteristic parameters. Since the $d$-dimensional singly spinning Kerr-AdS black hole system is a single-characteristic-parameter system, the critical point has an explicit form, i.e., Eq. (\ref{criticalpoint}). Thus, in order to obtain the critical point, we only need to derive the coefficients $\beta_{1}\sim\beta_{4}$. It is also worth noting that the critical point determined by this method is exact and obtained without any approximation.

Then we considered the generalized Maxwell equal-area laws in the extended phase space, and three critical conditions were obtained. Employing the condition (\ref{crit2}) and the unique form of the critical point (\ref{criticalpoint}), we obtained the analytical and exact critical points for the Kerr-AdS black hole for the first time.

We obtained the analytical critical point for the four-dimensional Kerr-AdS black hole. The coexistence curve and the phase diagram were also obtained. A fitting form of the coexistence curve in the $P$-$T$ phase diagram was presented, and it was found to be angular momentum $J$ independent. The $P$-$V$ and $T$-$V$ phase diagrams were also given. In particular, in the $T$-$V$ phase diagram, besides the small and large black hole phases, we gave two extra subphases, i.e., the superheated small black hole phase and supercooled large black hole phase.

When $d=5$, the critical phenomena are similar to the four-dimensional case, while when $d\geq6$ the reentrant phase transition occurs, and two exact critical points are presented. However, the small one does not participate in the phase transition. Through a simple calculation, we showed that the approximate critical point (\ref{appcrit}) obtained in the small-$J$ limit is accurate not only valid for small $J$, but also for large $J$. The reason is that the singly spinning Kerr-AdS black hole system is a single-characteristic-parameter system, and the form of the critical point is uniquely determined.

When $d\geq6$, the reentrant phase transitions appear. Based on the similar analysis for the critical point, we obtained the exact CRPs for $d=6$---9.

Finally, we suggested that the success in obtaining the analytical and exact critical points is mainly based on the fact that the $d$-dimensional singly spinning Kerr-AdS black hole is a single-characteristic-parameter system. Thus the form of the critical point is uniquely determined. For the multiply spinning black hole, this method fails, and the forms of the critical points cannot be easily determined using only the dimensional analysis. However, for the equal-spinning ($J_{1}=J_{2}= ...... = J$) Kerr-AdS black holes, this method is valid. We will consider the exact critical point and phase diagram for these black holes in future work.

\section*{Acknowledgements}
This work was supported by the National Natural Science Foundation of China (Grants No. 11205074, No. 11375075, and No. 11522541).

\end{document}